\documentclass[twocolumn,showpacs,preprintnumbers,amsmath,amssymb,pra,superscriptaddress]{revtex4}

\usepackage{graphicx}
\usepackage{dcolumn}
\usepackage{bm}

\def \figabbr{Figure\ }
\def \eqnabbr{Eq.\ }

\def \dBzdz{\partial B_z/\partial z}

\begin{document}

\title{Gravitationally enhanced depolarization of ultracold neutrons in magnetic field gradients, and implications for neutron electric dipole moment measurements
}
\author{P.G.\ Harris}\affiliation{Department of Physics and Astronomy, University of Sussex, Falmer, Brighton BN1 9QH, UK}
\author{J.M.\ Pendlebury}\affiliation{Department of Physics and Astronomy, University of Sussex, Falmer, Brighton BN1 9QH, UK}
\author{N.E. Devenish}\affiliation{Department of Physics and Astronomy, University of Sussex, Falmer, Brighton BN1 9QH, UK}

\date{\today}

\begin{abstract}
Trapped ultracold neutrons (UCN) have for many years been the mainstay of experiments to search for the electric dipole moment (EDM) of the neutron, a critical parameter in constraining scenarios of new physics beyond the Standard Model.  Because their energies are so low, UCN preferentially populate the lower region of their physical enclosure, and do not sample uniformly the ambient magnetic field throughout the storage volume.  This leads to a substantial increase in the rate of depolarization, as well as to shifts in the measured frequency of the stored neutrons.  Consequences for EDM measurements are discussed.

\end{abstract}

\maketitle

\section{Introduction}

Ultracold neutrons (UCN) are neutrons of extremely low energy, typically less than or of the order of 200 neV, which therefore have wavelengths that are long compared with the spacing between atomic nuclei in solids.  The surfaces of many materials then appear as a positive potential barrier (the so-called Fermi potential) from which these neutrons reflect.  This allows the storage of such neutrons in material bottles, typically for several minutes at a time, which in turn permits the study of their fundamental properties.  One such study is the ongoing search for the electric dipole moment (EDM) of the neutron, of which the most recent measurement was carried out at the Institut Laue-Langevin, Grenoble, by a collaboration led by the University of Sussex and the Rutherford Appleton Laboratory,\cite{baker06} using apparatus at room temperature (in contrast to its cryogenic successor, now under development).  

The internal volume of the neutron trap used in the room-temperature EDM experiment (RT-nEDM) was an upright cylinder 12 cm high, with quartz walls 37 cm in diameter and a roof and floor of aluminium coated with diamond-like carbon.  Crucial to the analysis of the experimental data was the fact that the UCN, being of very low energy, tended to populate preferentially the lower part of the storage volume, whereas the cohabiting mercury ($^{199}$Hg) magnetometer\cite{green98} filled the volume uniformly.  Any vertical magnetic-field gradient $\dBzdz$ applied to the volume would affect the two species differently, such that the gyromagnetic-ratio-corrected ratio of the neutron and mercury Larmor precession frequencies
\begin{equation}
\label{eqn:R}
R = \left| \frac{\nu_n}{\nu_{\rm Hg}}\cdot \frac{\gamma_{\rm Hg}}{\gamma_{n}} \right|
\end{equation}
would, to first order, be shifted by 
\begin{equation}\label{eqn:DeltaR}
\Delta R = \pm \Delta h \cdot \frac{\dBzdz}{B_{0_z}},
\end{equation}
where $\Delta h$ is the (always positive) difference in height between the centre of mass of the mercury and that of the UCN, and the $\pm$ sign depends upon the relative directions of $B_{0_z}$ and $\dBzdz$: $R$ increases (i.e.\ $\Delta R$ becomes more positive) as the absolute magnitude of the field at the bottom of the storage cell (sampled preferentially by the neutrons) increases relative to that at the top of the cell. 

The Larmor precession frequency of the UCN was measured by means of the Ramsey method of separated oscilliatory fields, for which a time $T$ = 130 s between the two r.f.\ pulses was used consistently.  During this period, the UCN would suffer some loss of their (transverse) polarization.  This study looks at some of the mechanisms and consequences of this so-called $T_2$ relaxation.  For the sake of example, all values of the various parameters used in modelling the phenomenon (storage cell size, Fermi potential, $B_{0_z}$ magnitude etc.)\ are those appropriate to the RT-nEDM experiment.  

\section{UCN density distributions}

It is convenient to refer to the energy of UCN in terms of the maximum height attainable within Earth's gravitational field.    Phase space arguments can be used to demonstrate\cite{golub_UCN_book,pendlebury94} that a population of trapped UCN each of energy $\epsilon$ has a density distribution with height $h$ of the form
\begin{equation}
\label{eqn:height_dist}
n(h) = \left(1-\frac{h}{\epsilon}\right)^{1/2}n(0).
\end{equation}
Integration and inversion of this function shows that the height distribution of such UCN within a storage cell of height $H$ may be generated from numbers $X$ distributed uniformly between 0 and 1 via the equation
\begin{equation}
\label{eqn:h_generator}
h = \epsilon\left[1-\left(1-kX\right)^{2/3}\right].
\end{equation}
The constant $k=1$ when $\epsilon<H$, and $k=1-\left(1-H/\epsilon\right)^{3/2}$ otherwise.  We note that the average height of the UCN within this population is
\begin{equation}
\label{eqn:average_height}
\left<h\right> = -\frac{\epsilon}{k}\left[0.6-k-0.6\left(1-k\right)^{5/3}\right].
\end{equation}

UCN may be generated by capturing the very low-energy tail of the Maxwell-Boltzmann distribution within a thermal source, or else by downscattering from e.g.\ liquid helium in a superthermal source.  In either case, the energy distribution tends to be close to
\begin{equation}
\label{eqn:energy_distn}
n(\epsilon)d\epsilon \propto \epsilon^{1/2}d\epsilon.
\end{equation}
By the time the UCN are stored, this distribution can change: for example, allowing the UCN to fall under gravity will shift the entire energy distribution upwards; or passage through a polarising foil can remove those of low energy.  The top of the energy distribution tends to have a fairly sharp cut-off, corresponding to the Fermi potential of the storage vessel.  In the case of RT-nEDM, the UCN rose under gravity after passage through a polarising foil, and the bottom of the storage cell was positioned such that those with just enough energy to pass through the foil would also have just enough energy to reach the cell.  Here,  therefore, the energy distribution is modelled with the simple function of \eqnabbr  \ref{eqn:energy_distn}, using the 93 cm equivalent height Fermi potential of the quartz walls of the vessel as the cutoff energy.  As above, integration and inversion yields a generating function
\begin{equation}
\label{eqn:E_generator}
\epsilon = \epsilon_{F}Y^{2/3},
\end{equation}
where $\epsilon_{F}$ is the (Fermi potential) cut-off energy, and the numbers $Y$ are distributed uniformly between 0 and 1.

\begin{figure} [ht]
\begin{center}
\resizebox{0.5\textwidth}{!}{\includegraphics{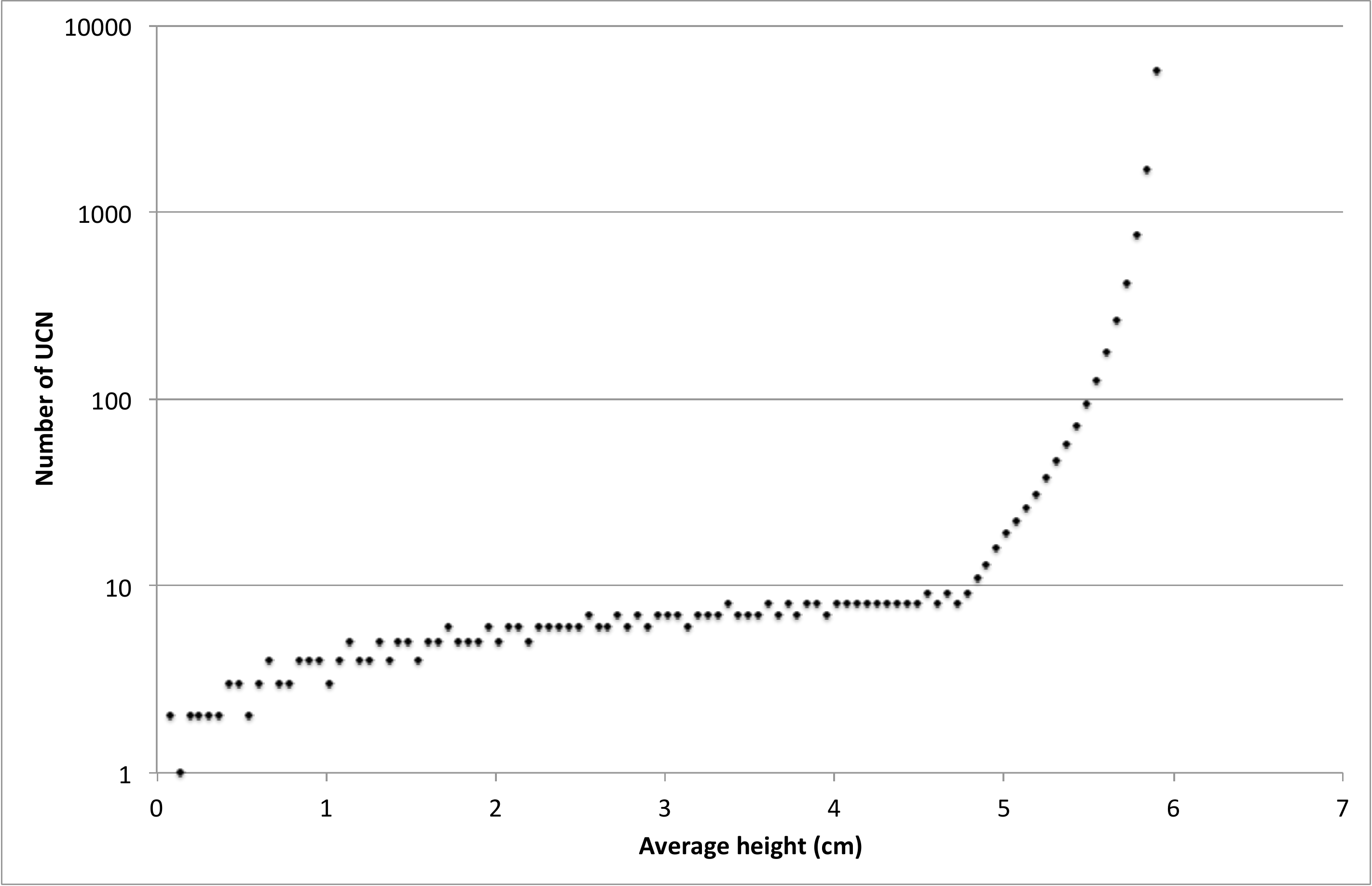}}
\end{center}
\caption{UCN density distribution as a function of height within a storage trap of height 12 cm and a Fermi potential of 93 cm height-equivalent. }
\label{fig:height_dist}
\end{figure}

 The distribution of average heights of a population of UCN with such an energy distribution is shown in \figabbr \ref{fig:height_dist}. The centre of mass of this modelled population of UCN is 3.0 mm below the centre of the storage trap, in good agreement with the 2.81 mm reported in Baker et al.\cite{baker06}.  Some 4.6\% of the UCN are not sufficiently energetic to reach the top of the trap. There is a small but extended tail, amounting to some 4.6\% of the total population, of neutrons that do not have sufficient energy to reach the lid of the bottle.  Over time, in a vertical magnetic-field gradient, two processes contributing to $T_2$ depolarisation come into play: (a) There is an energy dependence to the natural depolarisation rate in a magnetic-field gradient, because of the different rates at which the neutrons sample the measurement volume.  This will be referred to as the {\em intrinsic} component, and is modelled in this study by means of a simulation.  It is applicable even in the absence of a gravitational field.  (b) Under gravity, UCN at different average heights effectively sample different magnetic fields, and therefore on average precess at different rates.  This will be referred to as the {\em enhanced} component, and is here modelled by means of the analytic distributions described above.
 
\section{Simulation of UCN depolarisation in magnetic-field gradients}

Other studies have considered $T_1$ (longitudinal) and $T_2$ (transverse) relaxation rates of atoms in various configurations of electromagnetic fields and storage trap geometries.\cite{gamblin65, schearer65, cates88, cates88b, mcgregor90, schmid08}  An approach often adopted is that of the autocorrelation function, as outlined by McGregor.\cite{mcgregor90}  In this instance, however, the situation is complicated by the parabolic nature of the orbits of the UCN moving under gravity.  For this study, therefore, a Monte Carlo simulation has been developed, in which the UCN move in ballistic trajectories within the RT-nEDM cylindrical trap described above,  and their spins evolve classically according to the solution 
\begin{align*}
\vec{\sigma}(t) &= \left(\vec{\sigma}_0-\frac{\left(\vec{\sigma}_0\cdot \vec{B}\right)\vec{B}}{B^2}\right)\cos\left(\omega t\right) \\
                          &+ \frac{\vec{\sigma}_0\times\vec{B}}{B}\sin\left(\omega t\right) \\
                          &+ \frac{\left(\vec{\sigma}_0\cdot\vec{B}\right)
                          \vec{B}}{B^2}
\end{align*}
 of the equation of motion 
 \begin{equation*}
 \dot{\vec{\sigma}} = \gamma\vec{\sigma}\times\vec{B}
 \end{equation*}
of the spin $\vec{\sigma}$ in a magnetic field $\vec{B}$. A vertical holding field $B_{0_z}$ of 1 $\mu$T was applied.

\begin{figure} [ht]
\begin{center}
\resizebox{0.5\textwidth}{!}{\includegraphics{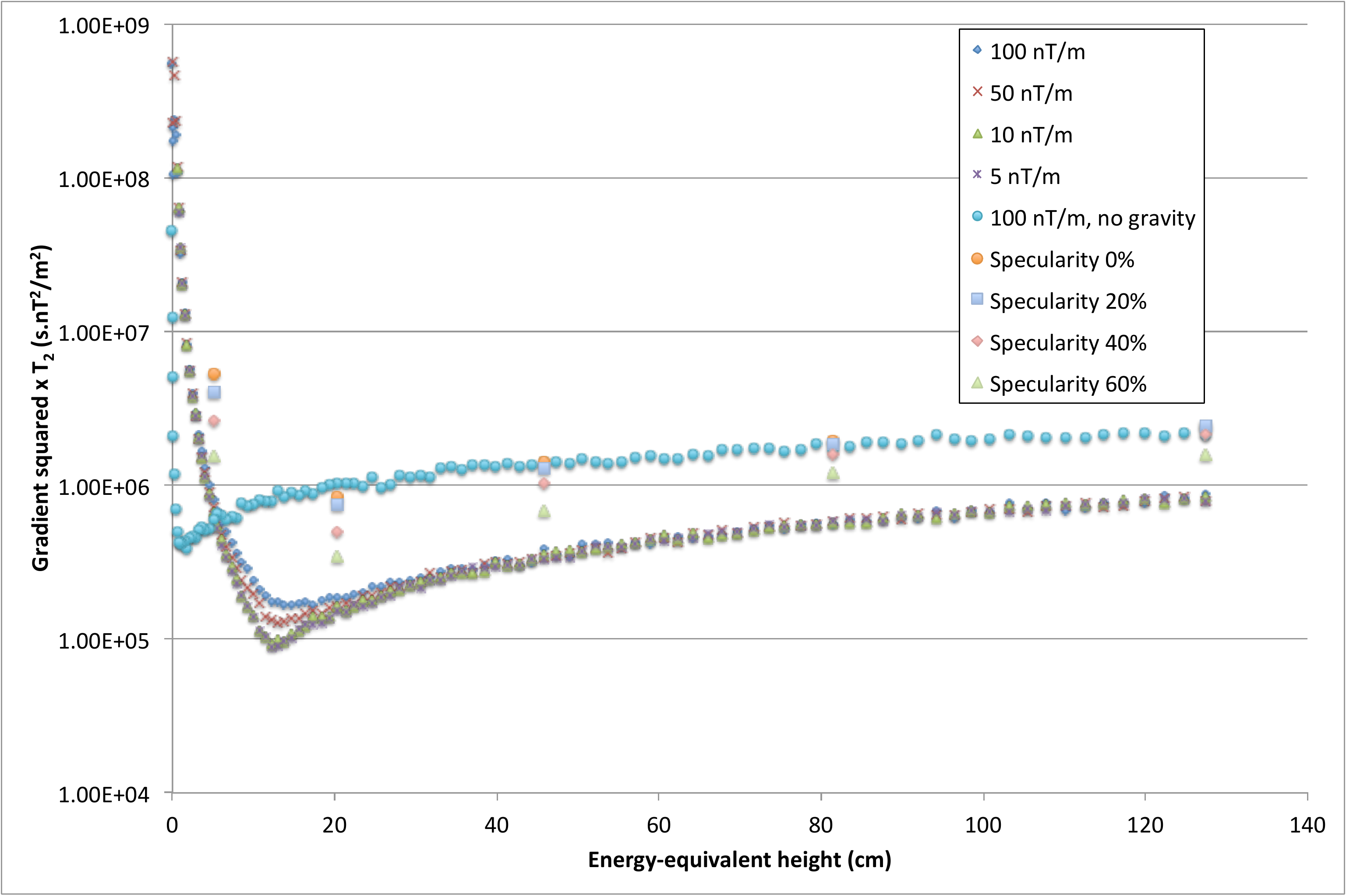}}
\end{center}
\caption{(Color online) $T_2$ multiplied by the square of the magnetic-field gradient as a function of UCN energy.  See text for details.}
\label{fig:T2_energy}
\end{figure}

Intuitively, since the polarisation is the ensemble average of projections $\cos(\delta \theta)\sim1-\delta\theta^2/2$, one can argue that the depolarisation rate should depend upon the variance of this quantity, and hence on the variance of $B$.  Thus, one expects that the (intrinsic) $T_2$ should depend inversely upon $\left(\partial B_z/\partial z\right)^2$.  \figabbr \ref{fig:T2_energy} shows  the values of $T_2\cdot(\partial B_z/\partial z)^2$, for a variety of different gradients, as a function of the UCN energy (represented by the maximum achievable height).  For convenience, the gradients are in nT/m, which is appropriate for the magnitudes of gradients to be expected in such experiments.  The scatter of the data points is representative of the uncertainty.  The dependence upon  $(\partial B_z/\partial z)^2$ provides an extremely good match across several orders of magnitude.  The minimum $T_2$ corresponds to the point at which the UCN just have sufficient energy to reach the roof of the trap.  For reference, the case in which gravity provides no influence is also shown.

In the simulations underlying \figabbr \ref{fig:T2_energy}, completely specular reflections were assumed to occur 80\% of the time.  In fact, $T_2$ also has a significant dependence upon the specularity because of the inclination of specular reflections to lead to individual UCN lingering in particular orbits within the trap rather  than sampling the volume uniformly.  Representative sample points are shown (based upon a field gradient of 100 nT/m) from the equivalent curves with specularities ranging from 0\% (perfectly diffuse reflections at each wall collision) to 60\%.


\section{Effect upon frequency and polarization}

In EDM measurements, the Larmor spin precession frequency is normally determined by the Ramsey method of separated oscillating fields.  This allows a precise measurement of the ensemble average difference in accumulated phase (per unit time) between the spins of the UCN population and the reference oscillator providing the spin-manipulating r.f.\ fields.  Over time, as different UCN sample different regions of the trap, the distribution of phases spreads out, and polarization is lost.  

\begin{figure} [ht]
\begin{center}
\resizebox{0.5\textwidth}{!}{\includegraphics{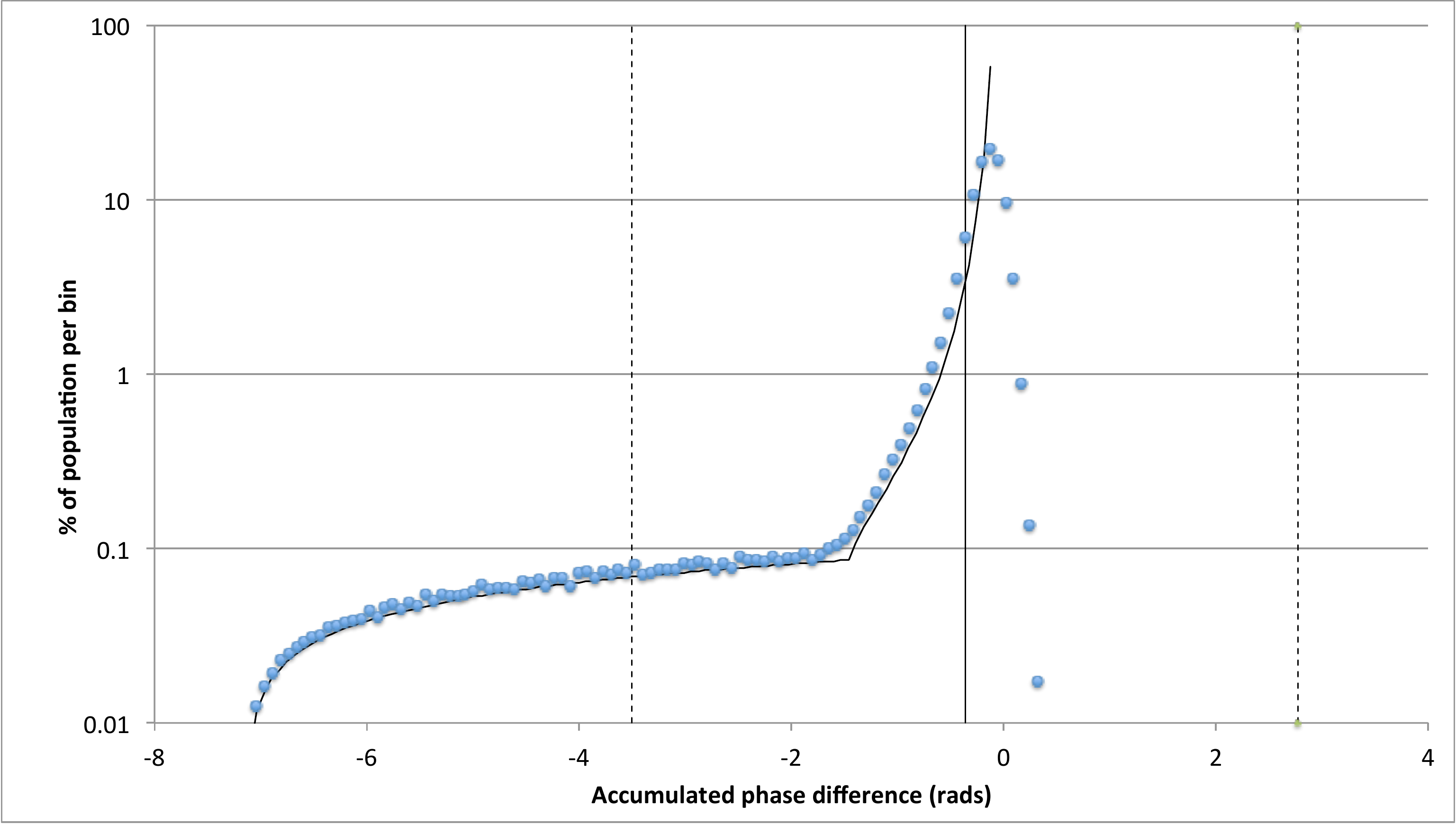}}
\end{center}
\caption{(Color online) Distribution of phase differences for a population of UCN in a $B$-field gradient of 5 nT/m after 130 s of Larmor precession.  See text for details.}
\label{fig:phase_dist}
\end{figure}

In the absence of a gravitational field, all of the neutrons would normally sample all regions of the trap with equal probability.  The distribution of accumulated phases would therefore be expected to be Gaussian, with the frequency determined by the phase at the peak of the distribution.  In the case of UCN, however, the distribution is skewed by the low-energy tail.  \figabbr \ref{fig:phase_dist} shows this distribution of phases for a measurement of 130 s duration (as used in RT-nEDM) in a magnetic-field gradient of 5 nT/m.  The numerical values of these phases are relative to that appropriate to the volume-averaged magnetic field, i.e. the field at the geometric centre of the trap.  The solid curve shows the distribution excluding the intrinsic contribution: the latter provides a relatively small additional spreading of the phases.   The reference phase $\hat{\phi}$, from which the frequency is determined, is given by
\begin{equation}
\hat{\phi} = \tan^{-1}\left( \frac{\left<\sin\phi\right>}{\left<\cos\phi\right>}   \right),
\end{equation}
where $\cos\phi$ and $\sin\phi$ are averaged over all of the individual phases $\phi$.  $\hat\phi$ is represented on the figure by the central (solid) vertical line, and it is clearly not at the peak of the distribution.   Also indicated (by dashed vertical lines) are the phases $\pm \pi$ away from the reference phase.

\begin{figure} [ht]
\begin{center}
\resizebox{0.5\textwidth}{!}{\includegraphics{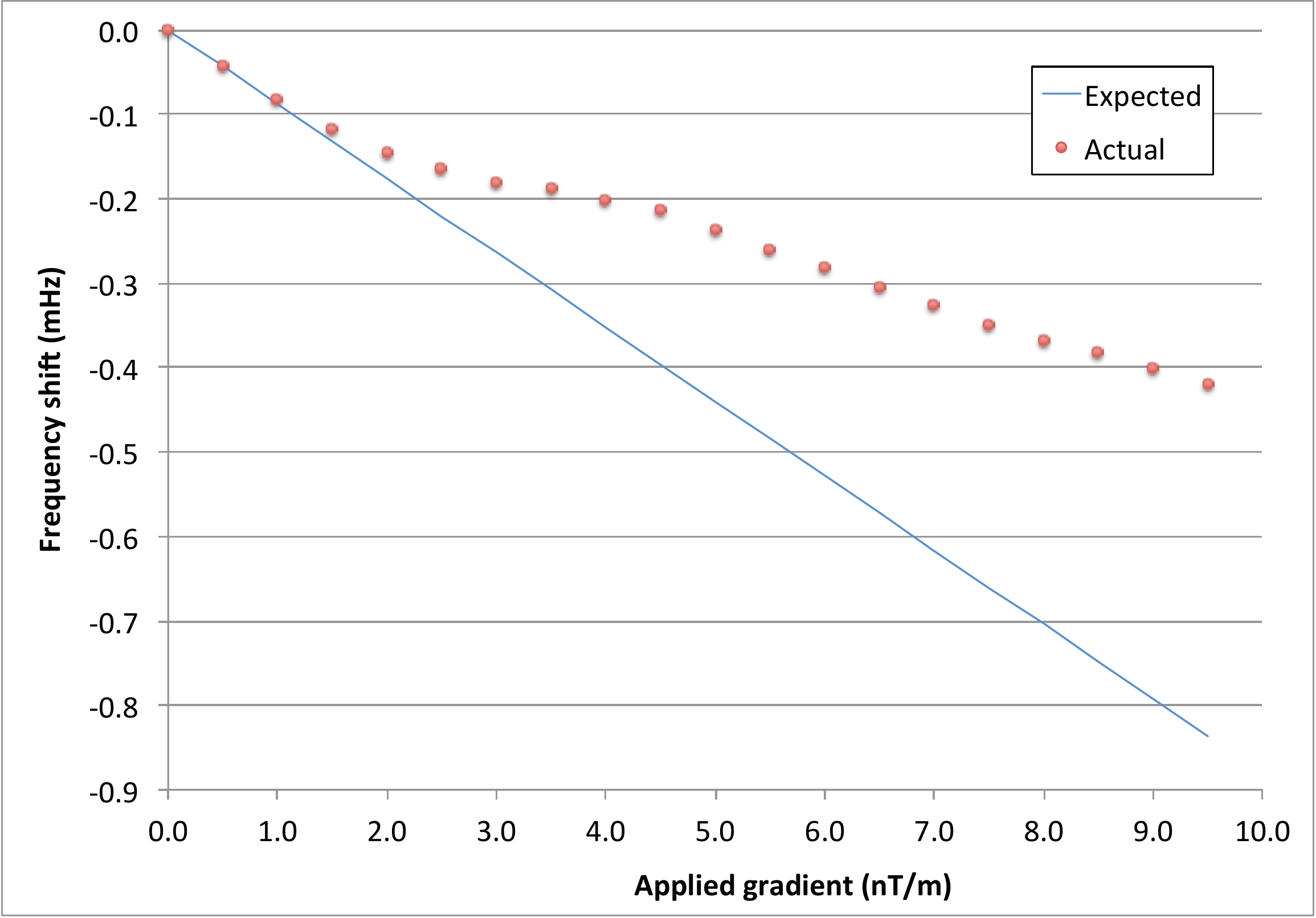}}
\end{center}
\caption{(Color online) Shift in measured frequency as a function of applied $B$-field gradient.  The straight line represents the value expected simply from the reduced height of the centre of mass of the neutrons.  See text for details.}
\label{fig:freq_shift}
\end{figure}

It will be noted that, as the phase distribution of \figabbr \ref{fig:phase_dist} spreads, the population within the low-energy tail passes beyond $\pi$ radians from the reference phase.  Since the Ramsey technique is sensitive only to phase modulo $2\pi$, these neutrons effectively reappear on the other side of the distribution, which pulls the reference phase back up towards the peak again, thus effectively {\em reducing} the frequency shift from the value (c.f. \eqnabbr \ref{eqn:DeltaR}) 
\begin{equation}
\Delta \nu = \gamma \Delta h \cdot \frac{\partial B_z}{\partial z}
\end{equation}
 that would naively be expected from the 3 mm height reduction of the centre of mass in combination with the applied magnetic field gradient. This effect is shown in \figabbr \ref{fig:freq_shift}.  
 
\begin{figure} [ht]
\begin{center}
\resizebox{0.5\textwidth}{!}{\includegraphics{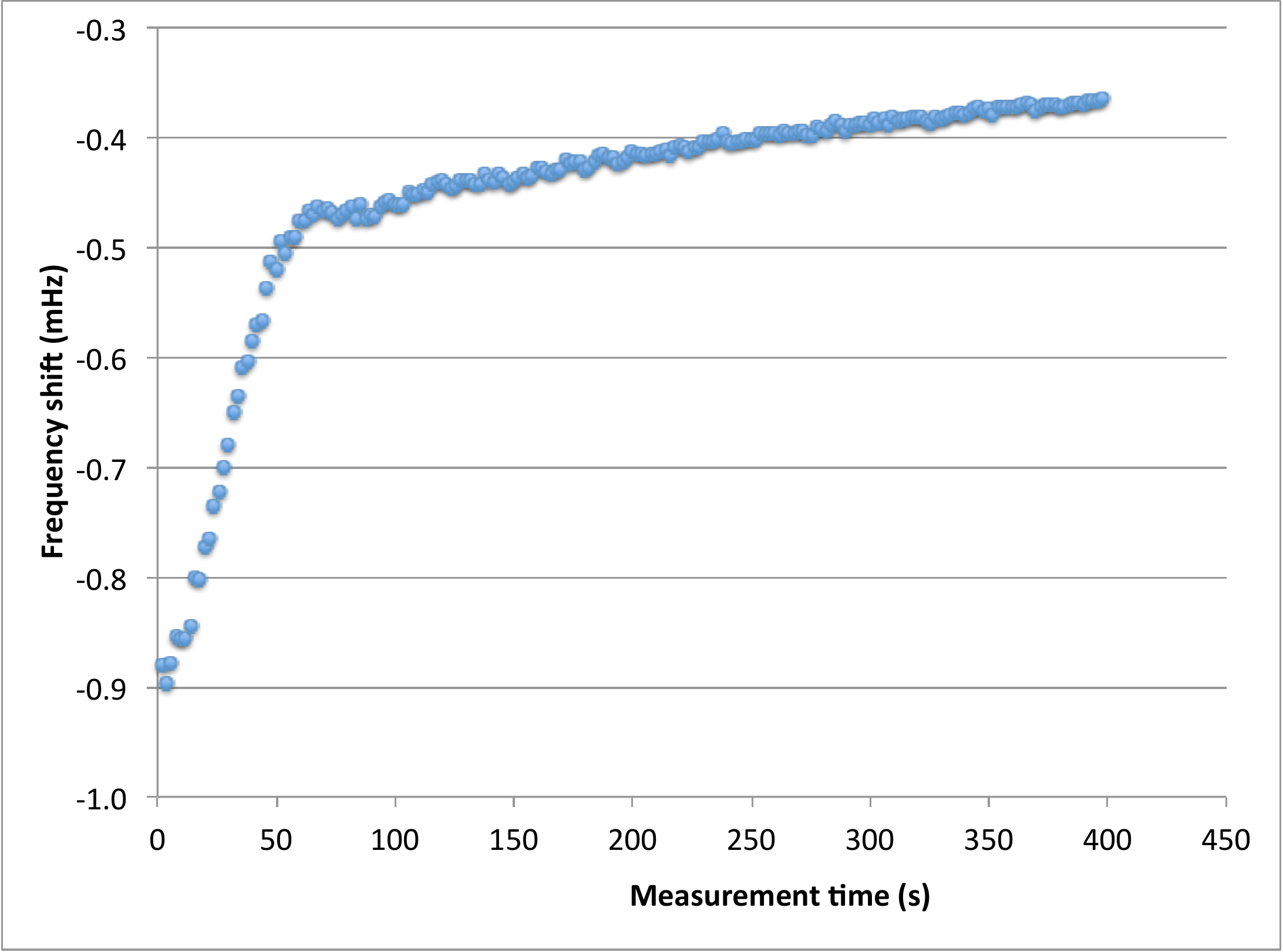}}
\end{center}
\caption{(Color online) Shift in measured frequency as a function of time, for a $B$-field gradient of 10 nT/m.  }
\label{fig:freq_vs_time}
\end{figure}

 The frequency shift also changes with time, as shown in \figabbr \ref{fig:freq_vs_time}.  This effect is likewise a direct consequence of the asymmetric nature of the distribution of phases in \figabbr \ref{fig:phase_dist}.  If the reference phase $\hat{\phi}$ were simply the average accumulated phase, the frequency would be constant for a given gradient.  The frequency change is initially fairly rapid, but then slows down as the tail of the distribution ``wraps around'' and starts to pull the reference phase back towards the peak.
 
There is an additional second-order frequency-shift effect.  Since the intrinsic contribution to the depolarization causes different parts of the energy spectrum to relax at different rates, and since depolarized UCN cannot contribute to the frequency measurement, the energy distribution of contributing UCN changes over time.  This in turn changes the effective centre of mass of the polarised UCN, and thus, via the applied field gradient, produces a second-order frequency shift.  The depolarization times $T_2$ in this scenario are however sufficiently long that this effect is negligible -- about two orders of magnitude smaller than the shifts that we have been considering thus far.

\begin{figure} [ht]
\begin{center}
\resizebox{0.5\textwidth}{!}{\includegraphics{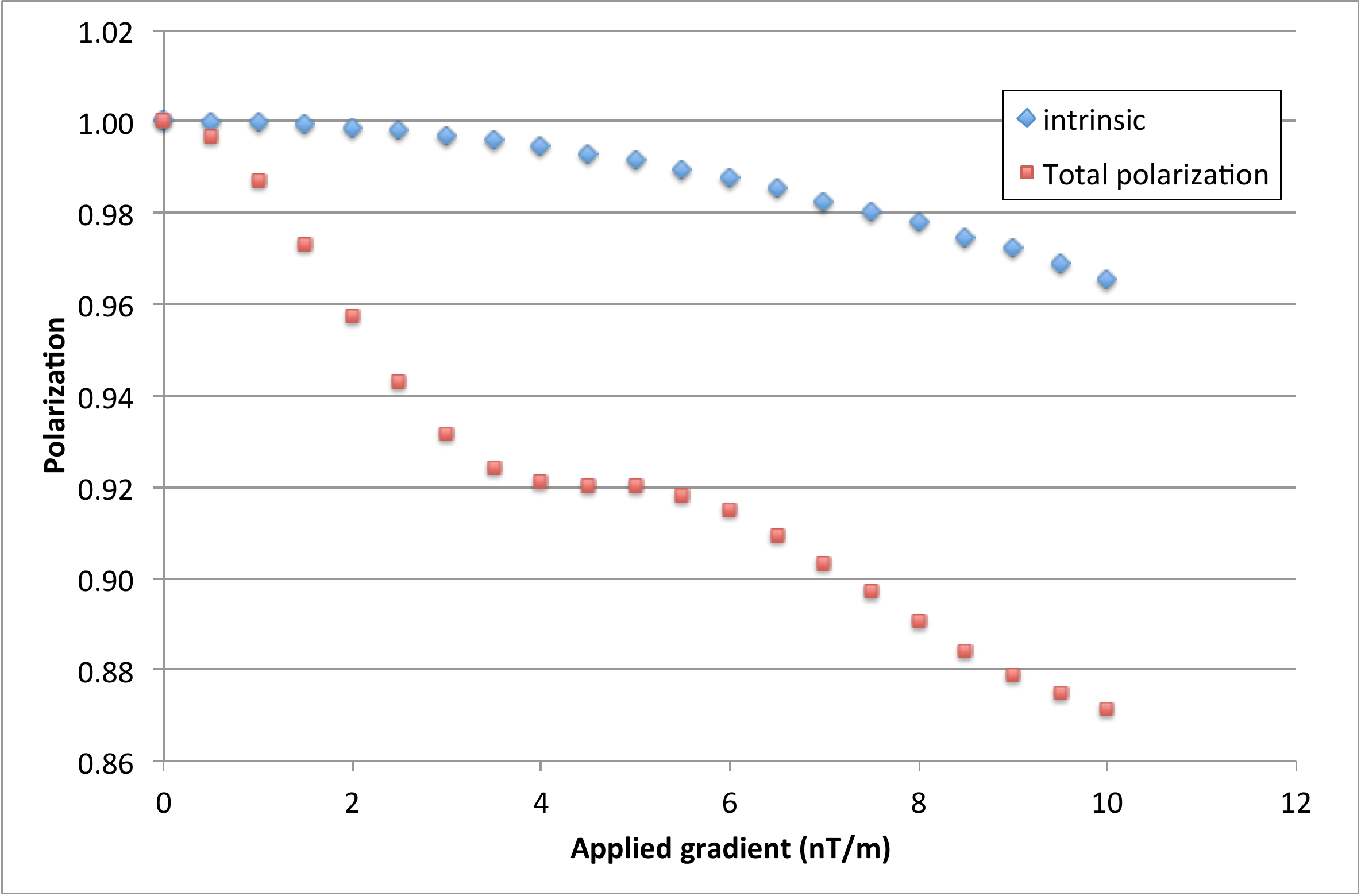}}
\end{center}
\caption{(Color online) Polarization as a function of applied magnetic-field gradient.  See text for details.}
\label{fig:alpha_peak}
\end{figure}

\begin{figure} [ht]
\begin{center}
\resizebox{0.5\textwidth}{!}{\includegraphics{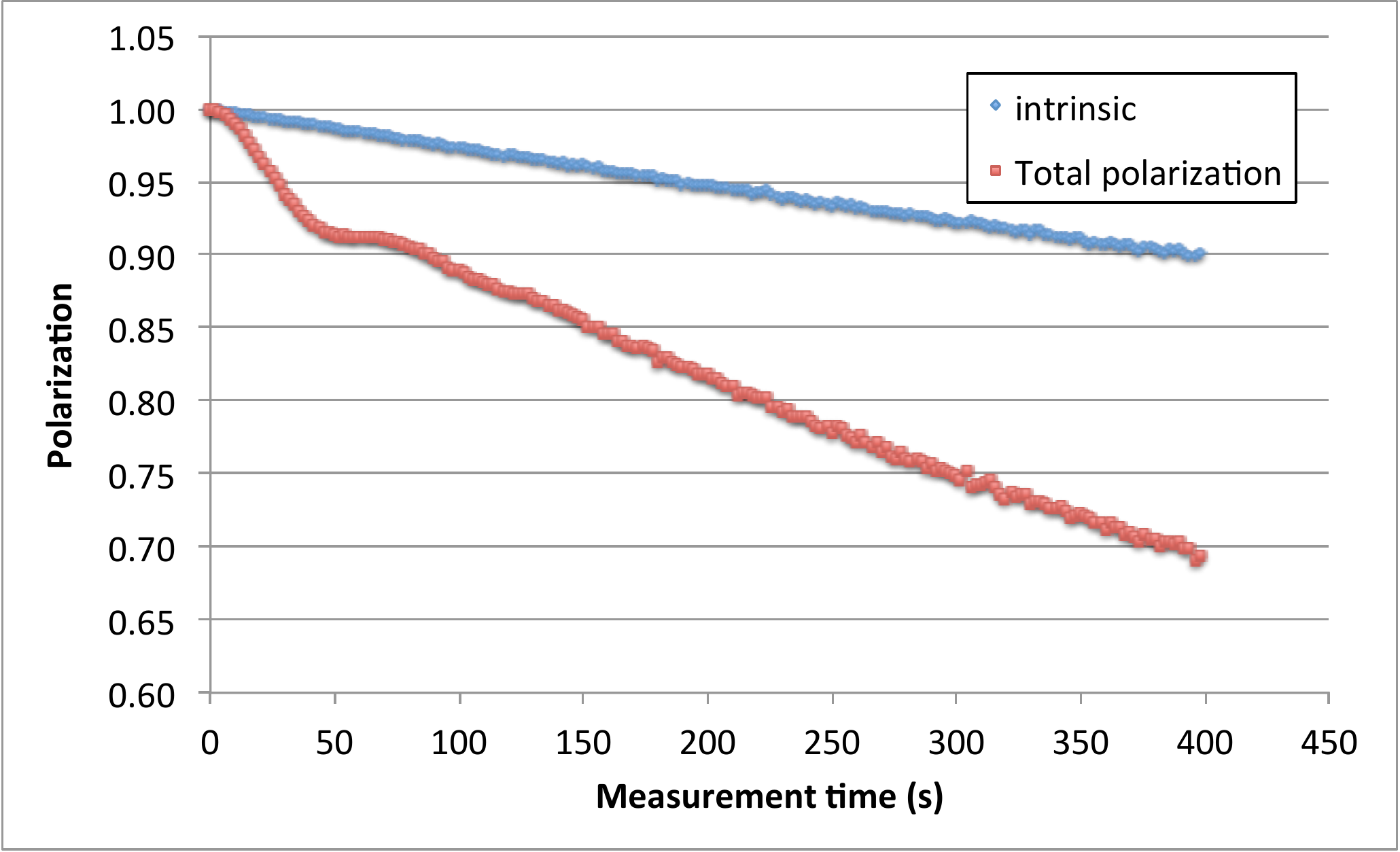}}
\end{center}
\caption{(Color online) Polarization $\alpha$ as a function of time, for a $B$-field gradient of 10 nT/m.  }
\label{fig:alpha_vs_time}
\end{figure}

The polarization $\alpha$ is the average projection onto the reference phase of all of the spin vectors. $\alpha$ as a function of the applied gradient, again for a measurement period of 130 s, is shown in  \figabbr \ref{fig:alpha_peak}. The intrinsic contribution to the depolarization is shown explicitly as a separate set of points.  The structure that is apparent at an applied gradient of 4-5 nT/m  is another effect of the tail of the distribution wrapping around and moving towards the peak from the other side: temporarily at least, it reduces the average spread of the distribution, and thus moderates the fall in polarization.  $\alpha$ as a function of time, for an applied gradient of 10 nT/m, is presented in \figabbr \ref{fig:alpha_vs_time}; once again, the intrinsic contribution (which falls off exponentially) is shown explicitly.

\section{Effect upon EDM systematic error calculations}

Whilst any depolarization will reduce the sensitivity of an EDM measurement, it does not of itself bias the result.  Likewise, a frequency shift that is independent of the electric field is not necessarily a cause for concern.  Effective changes in the velocity spectrum due to differential depolarization could in principle influence the geometric-phase (GP) contribution, although any such effect is liable to be completely negligible: in RT-nEDM, the GP contribution arising from the UCN was approximately fifty times smaller than that from the cohabiting mercury magnetometer.

A more significant concern is in the interpretation of GP-induced false-EDM signals.  These are proportional to the applied field gradient $\partial B_z/\partial z$.  In a real experiment such as RT-nEDM, this gradient can be most easily inferred from the ratio $R$ (see \eqnabbr \ref{eqn:R}) of the neutron frequency to the frequency of the cohabiting mercury magnetometer, which samples the volume uniformly.  The measured EDM signals as a function of this ratio are shown in   \figabbr 13 of \cite{pendlebury04}, and are fitted to the straight lines anticipated from \eqnabbr \ref{eqn:DeltaR} above. However, the frequency shifts due to the enhanced depolarization mean that the appropriate frequency ratio is more complex than this, and a function similar to that shown in \figabbr \ref{fig:freq_shift} is required instead.  Fitting to these lines should therefore be carried out with due care and attention, and only after careful modelling.  It is clearly far preferable to undertake EDM measurements in conditions of very low magnetic-field gradients.

\section{Conclusion}

UCN are of very low energy, and preferentially populate the lower regions of any trap within which they are contained.  It has been shown that these gravitational effects result in a significant enhancement of the $T_2$ relaxation of the UCN, and can also lead to shifts in the measured Larmor precession frequency.  Although there are potential impacts upon systematic-error calculations for EDM measurements, these are at a very manageable level; nonetheless, they underline the importance both of careful and precise modelling of the system, and also of keeping to an absolute minimum any magnetic-field gradients within the measurement apparatus.

\section*{Acknowledgments}
This work was supported in part by grant no.\ ST/K001329/1 from the UK Science and Technology Facilities Council.  

\bibliography{neutron_edm}

\end{document}